\newcommand{\beq}[1]{\begin{equation}\label{#1}}
\newcommand{\eeq}{\end{equation}}
\newcommand{\beqar}[1]{\begin{eqnarray}\label{#1}}
\newcommand{\eeqar}{\end{eqnarray}}

\documentstyle[12pt,dina4,preprint,aps,epsf]{revtex}
\begin{document}
%%%%%%%%%%%%%%%%%%%%%%%%%%%%%%%%%%%%%%%%%%%%%%%%%%%%%%%%%%
\begin{flushright}
UFTP 420/1996 
%\\  hep-ph/9607438
\end{flushright}
%%%%%%%%%%%%%%%%%%%%%%%%%%%%%%%%%%%%%%%%%%%%%%%%%%%%%%%%%%
\vskip0.7cm
\begin{center}
{\large \bf Analyzing the nucleon spin in weak interaction processes at HERA
energies}
\vskip0.7cm
{\em  \underline{M. Maul}, A. Sch\"afer}
\vskip0.4cm
{\em
        Institut f\"ur Theoretische Physik, J.~W.~Goethe Universit\"at Frankfurt\\
        $^{~}$60054 Frankfurt am Main, Germany\\
        e-mail: maul@th.physik.uni-frankfurt.de\\
}
\end{center}
We investigate the possibility to measure the spin content 
carried by the different quark flavors in a nucleon by means of 
polarized deep inelastic scattering with $W^\pm$ exchange
at HERA. Such measurements require a polarized proton beam.
The expected inclusive and semi inclusive asymmetries are sizable 
and for realistic luminosities
the expected statistical accuracies are good enough
to extract new and relevant information on the 
valence quark and the strange sea distributions. 
\newpage
\noindent
Presently the realizability of a polarized proton beam at DESY allowing for 
polarized electron-proton scattering at $s\approx 10^5 ({\rm\; GeV})^2$ is intensively 
discussed \cite{HERA95}. One of the novel possibilities opened by such experiments is 
the investigation of spin effects in
charged current (CC) reactions, where the outgoing neutrino
is detected by its missing momentum.
CC events probe a different combination
of spin-dependent  quark-distribution functions than
electromagnetic processes, thus allowing to extract information on the 
flavor decomposition of the polarized
quark distribution functions. These experiments would yield results 
similar to those expected from
semi inclusive fixed target experiments \cite{HERMES,FS89}. To put the full 
potential of such experiments into reality
it is necessary to have polarized electron as well as positron beams at HERA.
The most important property of CC reactions is that
they distinguish quark and antiquark flavors, allowing
to extract the 
polarized valence quark distributions $ \Delta u_V$, $\Delta d_V$ as well as
the strange ones $\Delta s, \Delta \bar s$.
The latter two are usually assumed to be identical. If a difference between 
them were observed this could be related to the
fact that the virtual coupling of 
$p$ to e.g. ($K^+$ + baryon) is stronger than to 
($K^-$+baryon).
Including deep inelastic weak interaction processes, but neglecting lepton and
quark masses the hadronic scattering tensor can be decomposed  
into eight structure functions \cite{Ba79}:
\beqar{10}
\frac{1}{2m_N} W_{\mu \nu} &=& 
             - \frac{g_{\mu \nu}}{m_N} F_1(x,Q^2) 
             + \frac{p_\mu p_\nu}{m_N p \cdot q} F_2(x,Q^2)
\nonumber \\
&&             + \frac{i \epsilon_{\mu \nu \alpha \beta}}{2p \cdot q}
               \left[ \frac{p^\alpha q^\beta}{m_N} F_3(x,Q^2)
                      + 2 q^\alpha S^\beta  g_1(x,Q^2)
                      - 4xp^\alpha S^\beta  g_2(x,Q^2) \right]
\nonumber \\ && - \frac{p_\mu S_\nu + S_\mu p_\nu}{ 2p \cdot q}  g_3(x,Q^2)
              + \frac{S \cdot q}{(p \cdot q)^2} p_\mu p_\nu g_4(x,Q^2) 
              + \frac{S \cdot q}{p \cdot q} g_{\mu \nu} g_5(x,Q^2) \quad.
\eeqar          
In the quark parton model $(g_4 = 0)$ the  structure functions $g_3$ 
and $g_5$ fulfill in analogy to the Callan Gross relation
$2xg_5(x) = g_3(x)$ \cite{Di72,An94,BK96,Ma96}. 
Contracted with the leptonic scattering tensor the differential
cross section reads
for the CC reactions with longitudinally polarized targets
\beq{20}
\frac{d^2 \sigma}{dx dQ^2} =
\frac{G_F^2 M_W^4}{4 \pi} 
\Bigg( [ a F_1 - \lambda b F_3 ]  
+ [- \lambda 2b g_1 + a g_5] \delta \Bigg)  
\frac{1}{(Q^2 + M_W^2)^2} \quad,
\eeq
where $G_F$ is Fermi's constant and $M_W$ the W-Boson mass.
$\lambda$ is $-1$ for $e^-$ and $+1$ for $e^+$,
and $\delta$ denotes the longitudinal polarization of the nucleon 
antiparallel ($+1$) or parallel ($-1$) to the polarization
of the lepton.
The $y$ dependence is absorbed in the two constants $a$ and $b$ 
\beq{30}
a := 2(y^2 - 2y + 2) ; \qquad b := y(2-y) \quad.
\eeq
The asymmetry is defined by 
\beq{40}
A := \frac{d\sigma_{\uparrow \downarrow} - d\sigma_{\uparrow \uparrow}}
          {d\sigma_{\uparrow \downarrow} + d\sigma_{\uparrow \uparrow}} 
\quad,
\eeq
which leads to \cite{Ka96}
\beq{50}
A^{W^-} = \frac{2b g_1 + a g_5}{ a F_1 + b F_3 }
\; ; \qquad    
A^{W^+} = \frac{- 2b g_1 + a g_5}{ a F_1 -  b F_3 }
\quad.
\eeq
The ratio of the factors $a$ and $b$ determines the
contribution of the different structure functions to  the asymmetry.
For small $y$ values, for which the experimental statistics is best, $a$ is
much larger than $b$.
\newline
In the framework of the quark parton model the eight 
combinations of distribution functions tested by CC asymmetries
read \cite{An94}:
\beqar{60}
F_1^{W-} = u + c + \overline d + \overline s \; &,&\quad
F_1^{W+} = d + s + \overline u + \overline c \quad,
\nonumber \\
F_3^{W-} =  2( u + c - \overline d - \overline s) \; &,& \quad
F_3^{W+} = 2(d + s - \overline u - \overline c) \quad,
\nonumber \\
g_1^{W-} =\Delta  u + \Delta c + 
\Delta \overline d + \Delta \overline s \; &,& \quad 
g_1^{W+} =\Delta  d + \Delta s + 
\Delta \overline u + \Delta \overline c \quad,
\nonumber \\
g_5^{W-} =  \Delta u + \Delta c - 
\Delta \overline d - \Delta \overline s \; &,& \quad
g_5^{W+} = \Delta d + \Delta s - 
\Delta \overline u - \Delta \overline c \quad.
\eeqar
Rewriting (\ref{50}) in terms of the parton distributions we get for the
inclusive asymmetries:
\beqar{100}
A^{W^-}&=&\frac{\Delta u + \Delta c -[(y-1)^2][\Delta \overline s + \Delta \overline d]}
            { u  + c    +[(y-1)^2][       \overline s +        \overline d]}
\quad,
\nonumber \\
A^{W^+}&=&\frac{[(y-1)^2][\Delta  s + \Delta  d] - \Delta \overline u - \Delta \overline c}
               {[(y-1)^2][        s +         d] +        \overline u +  \overline c  }   
\quad.
\eeqar
In fig.~1 we show the expected asymmetries for HERA energies 
($E_l = 27.6\;{\rm\;  GeV}, \quad E_n = 820\;{\rm\; GeV} $)  with 
integrated cross sections in the $Q^2$ range $600-1000{\rm\;( GeV)}^2$ 
and $1000 {\rm\;( GeV)}^2 - {\rm  maximum}$.
We split the data into these two $Q^2$ ranges because it might not
be obvious that the detection of missing momentum in the range
$\sqrt{600}{\rm\; GeV} - \sqrt{1000} {\rm\; GeV}$ is absolutely reliable.  
(Usually already $Q^2 \geq 600 {\rm\; (GeV)}^2$ is regarded as save). Fig. 1 
shows that the low-$Q^2$ data do not dominate the
statistics. So even if one would be very 
restrictive on the criteria for identifying CC events our conclusions would 
be hardly affected. Indeed, the situation has recently been improved. 
At present in H1 it is possible to control CC events with
$p_t > 15 {\rm\; GeV}$, i.e. $Q^2 > 225 {\rm\; (GeV)}^2$, so one could easily
add another bin $[400-600]{\rm\; (GeV)}^2$ \cite{Feltesse}.
For the parton distributions we use the set "standard scenario, LO" 
given in \cite{GRSV95}. 
The error bars are calculated according to
\beq{80}
\Delta A = \sqrt{\frac{ 1-A^2}{ 2 {\cal L} \Delta x d \sigma/dx  }}
\quad,
\eeq
where $\cal{L}$ denotes the luminosity, and 
$d\sigma$ is the unpolarized cross section. We assumed here the 
luminosity to be $100 {\rm \;pb}^{-1}$ and 100$\%$ polarization.
This is identical to $200 {\rm \;pb}^{-1}$ and the beam polarization 
product $P_e  P_p = 0.5$.
(Such a luminosity is planned if proton 
polarization is realized at HERA. For the 
far future, options to raise it to $1000 {\rm \; pb}^{-1}$ are discussed.)
The remarkable features of our results are that the asymmetries 
are sizable and of considerable
extent in the range  $x \in [ 0.01,1]$, and that even for $100 {\rm\; pb}^{-1}$
the expected statistical accuracy 
for both $W^+$ and $W^-$ exchange is rather good. 
\newline
\newline
In fig.~2 we compare the contribution of the sea quarks and  valence 
quarks to the asymmetries using two different parton distribution sets.
One is again the LO standard scenario \cite{GRSV95}, the other one is 
LO Gluon C \cite{GS95}. The main distinction between
the two sets is the fact that the polarized valence quark distribution
given in \cite{GRSV95} is nearly proportional to the unpolarized valence
quark contribution, whereas in \cite{GS95} the fit shows larger deviations.
The total asymmetry is in \cite{GS95} even larger than in \cite{GRSV95}.
Obviously the expected accuracy should allow to distinguish the 
two scenarios.
Besides these single $W$ boson asymmetries one can form four combinations
which are determined by pure valence contributions and
by valence + sea contributions:
\beq{81}
A_{(\pm)_1 (\pm)_2}  := 
\frac{(d\sigma^{W^-}_{\uparrow \downarrow} - d\sigma^{W^-}_{\uparrow \uparrow})
 (\pm)_1  (d\sigma^{W^+}_{\uparrow \downarrow} - d\sigma^{W^+}_{\uparrow \uparrow})}
     {(d\sigma^{W^-}_{\uparrow \downarrow} + d\sigma^{W^-}_{\uparrow \uparrow})
 (\pm)_2  (d\sigma^{W^+}_{\uparrow \downarrow} + d\sigma^{W^+}_{\uparrow \uparrow})}
\quad.
\eeq
Here the first $\pm$ accounts for the numerator and the 
second for the denominator. With this definition we write: 
\beqar{82}
A_{++}
&=& \frac{\Delta u_V + [(y-1)^2] \Delta d_V}
         {u_T + c_T + [(y-1)^2][ s_T + d_T]} \quad,
\nonumber \\
A_{--}
&=& \frac{ \Delta u_T + \Delta c_T -
[(y-1)^2] [ \Delta s_T  + \Delta d_T]}
{ u_V - [(y-1)^2] d_V} \quad,
\nonumber \\
A_{-+}
&=& \frac{ \Delta u_T  + \Delta c_T -
[(y-1)^2] [ \Delta s_T +  \Delta d_T]}
{u_T + c_T + [(y-1)^2][ s_T + d_T]} \quad,
\nonumber \\
A_{+-}
&=&  \frac{\Delta u_V + [(y-1)^2] \Delta d_V}
{ u_V - [(y-1)^2]   d_V} \quad.
\eeqar 
Here we defined for the quark distribution functions $q_f$
\beqar{83}
\Delta q_f &=& q_{f, \uparrow \downarrow} - q_{f, \uparrow \uparrow} \quad, 
\nonumber \\
q_{f,V}    &=& q_f - \overline q_f \quad,
\nonumber \\
q_{f,T}    &=& q_f + \overline q_f \quad.
\eeqar
While the inclusive asymmetries are mainly sensitive to the
u- and d- quark flavor it is possible to extract information on
the strange and antistrange sea by semi inclusive CC current 
measurements.
As usual, we define the  fragmentation function $D_f^H(z)$ which
describes the probability that a quark with flavor $f$ fragments into
a hadron $H$ which carries the energy fraction $z$ of the whole energy 
transferred to the nucleon. 
We want to discuss semi inclusive $\pi$ and $D$ production.
As suggested in \cite{GVGMS93} we make the assumptions.
\beqar{90}
D^{\pi}_1(z)    & =&  D_u^{\pi^+}(z)  
         =  D_d^{\pi^-}(z)  
         =  D_{\overline d} ^{\pi^+}(z)  
         =  D_{\overline u} ^{\pi^-}(z)  \quad,
\nonumber \\
D^{\pi}_2(z)    & =&  D_d^{\pi^+}(z)
                   =  D_u^{\pi^-}(z) 
                   =  D_{\overline u} ^{\pi^+}(z)  
                   =  D_{\overline d} ^{\pi^-}(z)    
                   =  D_s^{\pi^+}(z)
                   =  D_s^{\pi^-}(z) 
\nonumber \\
         &  = & D_{\overline s} ^{\pi^+}(z)  
            =  D_{\overline s} ^{\pi^-}(z)   
            =  D_c^{\pi^+}(z)
            =  D_c^{\pi^-}(z)
            =  D_{\overline c} ^{\pi^+}(z)
            =  D_{\overline c} ^{\pi^-}(z) \quad.
\eeqar
In case of the fragmentation into $D$-mesons we have
\begin{equation}
D^{D}_1(z)      =  D_c^{D}(z)
                =     D_{\overline c}^{D}(z) \quad.
\end{equation}
The fragmentation of any other flavour into a
$D$ meson is strongly suppressed. The only relevant contributions come
from $s + W^+ \to D^+,{\rm or}\; D^0$. Also we set
\begin{equation}
\int_{0.2}^1 D^{D}_1(z) dz \approx 1 
\end{equation}
because each c-quark fragments in at least one charmed hadron.   
Furthermore the fragmentation functions are assumed to be spin independent.
The form of the fragmentation functions is taken from \cite{EMC89}.
For electromagnetic interactions the semi inclusive pion asymmetries are
the most interesting quantities because of the large counting rates.
For CC asymmetries this is not the case, because the tested asymmetries are
basically the same as for purely inclusive reactions.
(This is due to the fact that for up and down 
quarks weak isospin and strong isospin coincide.)
For the single $W$ asymmetries we get
\beqar{110}
A^{W-,\pi^-} &:=& \frac{ d \sigma_{\uparrow \downarrow} ^{\pi^-} - 
                       d \sigma_{\uparrow \uparrow}    ^{\pi^-}} 
                     { d \sigma_{\uparrow \downarrow}  ^{\pi^-} +
                       d \sigma_{\uparrow \uparrow}    ^{\pi^-}} 
\nonumber \\
&=& \frac{ \Delta u + \Delta c 
         -(y-1)^2  
         (        \Delta \overline d +   \Delta \overline s)   
         + \eta(z) (\Delta c - (y-1)^2 \Delta \overline s)} 
         {        u  +      c 
         +(y-1)^2
         (                 \overline d +         \overline s)   
         + \eta(z) (         c + (y-1)^2         \overline s)}
\nonumber \\
 &\approx&  A^{W^-}\left(  1 + \eta(z) \left[ 
\left( \frac{\Delta c}{G_{W^-}}  
      -\frac{       c}{F_{W^-}}\right)
-(y-1)^2
\left( \frac{\Delta \overline s}{G_{W^-}} 
      +\frac{       \overline s}{F_{W^-}}\right)               
\right] \right) \quad,
\eeqar
with
\beqar{120}
t &=& \tan^2 \theta_C \quad,
\nonumber \\
\eta(z) &=& (D^{\pi}_2(z) - D^{\pi}_1(z)) \frac{1-t^2}{D^{\pi}_1(z) + t^2 D^{\pi}_2(z)}
\quad,
\nonumber \\
F_{W^-} &=&    u  +  c +(y-1)^2 (\overline d + \overline s) \quad,
\nonumber \\
G_{W^-} &=&  \Delta u + \Delta c
         -(y-1)^2 (\Delta \overline d +   \Delta \overline s) \quad.
\eeqar
Analogously, in the case of $W^+$ exchange we obtain
\beqar{130}
A^{W^+,\pi^+} &:=&\frac{ d \sigma_{\uparrow \downarrow}  ^{W^+,\pi^+} - 
                         d \sigma_{\uparrow \uparrow}    ^{W^+,\pi^+}} 
                       { d \sigma_{\uparrow \downarrow}  ^{W^+,\pi^+} +
                         d \sigma_{\uparrow \uparrow}    ^{W^+,\pi^+}} 
\nonumber \\
&=& \frac{ -\Delta \overline u - \Delta \overline c 
         +(y-1)^2  
         (        \Delta d +   \Delta s)   
         + \eta(z) (-\Delta \overline c + (y-1)^2 \Delta s)} 
         {         \overline u  + \overline c 
         +(y-1)^2
         (                   d  +           s)   
         + \eta(z) (   \overline  c + (y-1)^2   s)}
\nonumber \\
 &\approx&  A^{W^+}\left(  1 - \eta(z) \left[ 
\left( \frac{\Delta \overline c}{G_{W^+}}  
      +\frac{       \overline c}{F_{W^+}}\right)
+(y-1)^2
\left(-\frac{\Delta s}{G_{W^+}} 
      +\frac{       s}{F_{W^+}}\right)               
\right] \right) \quad,
\eeqar
with 
\beqar{140}
F_{W^+} &=&      \overline u + \overline c
         +(y-1)^2 (d + s) \quad,
\nonumber \\
G_{W^+} &=&  - \Delta \overline u - \Delta \overline c
         +(y-1)^2 (\Delta d +   \Delta s) \quad.
\eeqar
For our numerical simulations we integrated
the fragmentation functions  over $0.2 < z < 1$,
which proves to be a good compromise between high statistics and
a good ratio of leading to non-leading particles.
For the parametrization of the fragmentation functions we used 
\beq{160}
D^{\pi}_1(z) = 0.665 \frac{(1-z)^{1.75}}{z} \;,\quad
D^{\pi}_2(z) = \frac{1-z}{1+z} D^{\pi}_1(z) \quad,
\eeq
which is an improved fit along the line of \cite{Au85}
to the newer data from \cite{EMC89}.
Fig.~3 shows the differences $A^{W^+,\pi^+}-A^{W^+}$ and
$A^{W^-,\pi^-}-A^{W^-}$. In those figures we use the set 
LO "standard scenario" from \cite{GRSV95} which does not contain
any contribution from the charm quarks. 
As shown in fig.~1, the anticipated accuracy for HERA CC events is of the 
order of per cent while the signal in fig.~3 is of the
order of per mille. Thus the
differences $A^{W^+,\pi^+}-A^{W^+}$, and
$A^{W^-,\pi^-}-A^{W^-}$ are too small to be useful as a new signal
for the nucleon spin decomposition. However, it offers  an urgently needed
opportunity to test one of the basic underlying assumptions,
crucial for all semi inclusive signals, namely the
spin independence of the fragmentation functions. Obviously this 
assumption could be tested at the percent level with CC pion asymmetries.
\newline
The situation for  
charm production is much more favourable. The resulting 
formulas for single asymmetries read
\beqar{190}
A_{W^-}^D 
:= \frac{   d \sigma^{\uparrow \downarrow}_{W^-,D} 
                        - d \sigma^{\uparrow \uparrow}_{W^-,D}}
                    {     d \sigma^{\uparrow \downarrow}_{W^-,D}
                        + d \sigma^{\uparrow \uparrow}_{W^-,D}}
           = -\frac{\Delta \overline s  + t^2  \Delta \overline d}
                   {       \overline s  + t^2         \overline d}
\quad,
\nonumber \\
A_{W^+}^D 
:= \frac{   d \sigma^{\uparrow \downarrow}_{W^+,D}  
                        - d \sigma^{\uparrow \uparrow}_{W^+,D}}
                    {     d \sigma^{\uparrow \downarrow}_{W^+,D}
                        + d \sigma^{\uparrow \uparrow}_{W^+,D}}
           =  \frac{\Delta           s  + t^2  \Delta           d}              
                   {                 s  + t^2                   d}
\quad.
\eeqar 
These asymmetries are obviously a direct signal for the strange and 
antistrange polarization and fig.~4 shows that the anticipated
statistical accuracy for ${\cal L}=1000 {\rm \; pb}^{-1}$ should allow 
for a very precise experimental determination.
Even for ${\cal L}=100 {\rm \; pb}^{-1}$ the signal still allows to
estimate the size of $\Delta s$ and $\Delta \bar s$.
For the high luminosity ${\cal L}=1000 {\rm \; pb}^{-1}$ it even appears
feasible to detect possible differences 
between $\Delta s (x)$ and  $\Delta \bar s(x)$ which could be 
induced by the fact that a proton emits more readily virtual
${K^+}'s$ than ${K^-}'s$.
We demonstrated that it is possible to measure at 
a polarized HERA with good accuracy
both inclusive and semi inclusive CC spin asymmetries.
Already for $100{\rm \;pb}^{-1}$ one should obtain a good determination
of $\Delta u_V$ and $\Delta d_V$ and a rough estimate for $\Delta s$
and $\Delta \bar s$, the latter for semi inclusive $D$ production.
For ${\cal L}=1000 {\rm \;pb}^{-1}$ semi inclusive $D$ production is an 
especially interesting observable . 
We conclude also, that semi inclusive pion asymmetries offer a very good
opportunity to test the
assumed absence of spin effects in string fragmentation.
\newline
\newline
{\bf Acknowledgement}
\newline
This work grew out of the HERA-Workshop program 1995/96 at DESY.
The discussion of polarized CC events was actually started by 
J.~Kalinowski, M.~Anselmino, and P.~Gambino \cite{Ka96,Ka96a}. 
For the inclusive single
$W$ boson asymmetries which both groups calculated the results agree.
This work has been 
supported by DESY, GSI, and BMBF.  A.S. also thanks
the MPI f\"ur Kernphysik in Heidelberg for support.

\newpage
{\LARGE  Figure Captions}
\newline
\newline
\newline
{  Figure~1:}
Asymmetries  $A^{W^+}$ and $A^{W^-}$ in the $Q^2$ range 600-1000 
${\rm\;( GeV)}^2$,
and 1000- max ${\rm\; (GeV)}^2$
for the luminosity of 100 ${\rm \;pb}^{-1}$ using the parton distributions
of \cite{GRSV95}.
\newline
\newline
{  Figure~2:}
Asymmetries $A^{W^+}$ and $A^{W^-}$ 
in the $Q^2$ range 600-max ${\rm\; (GeV)}^2$,
comparison between valence and sea quark contributions using set
``Gluon C, leading order ''\cite{GS95} and
``standard scenario, leading order '' \cite{GRSV95}.
\newline
\newline
{  Figure~3}   
Asymmetry differences
$A_{W+}^{\pi+} - A_{W+}$ and $A_{W-}^{\pi-} - A_{W-}$
in the $Q^2$ range 600-max ${\rm\; (GeV)}^2$
using ``standard scenario, leading order''\cite{GRSV95}.
\newline
\newline
{  Figure~4}
Charmed asymmetries 
$A_{W+}^{D}$ and $A_{W-}^{D}$
in the $Q^2$ range 600-max ${\rm\; (GeV)}^2$
using ``standard scenario, leading order'' \cite{GRSV95}.
The error bars are given for the luminosity of 1000 ${\rm \;pb}^{-1}$.
\newpage
\begin{center} 
\leavevmode
\end{center}
\begin{center}
\leavevmode
\epsfxsize=11.0cm 
\epsfbox{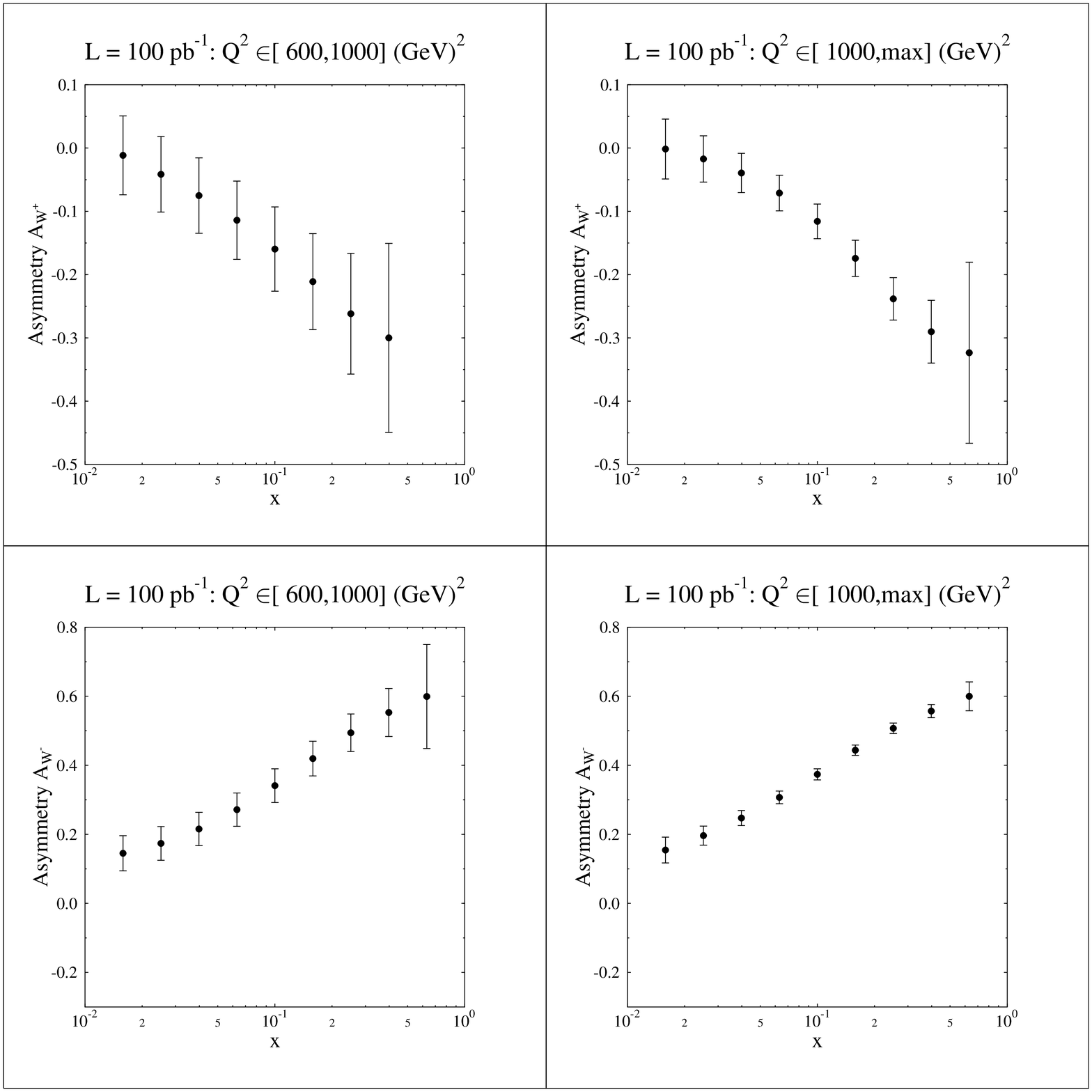}
\newline
\newline
\end{center}  
\begin{center}
{\bf Figure 1}
\end{center}
\begin{center}
\leavevmode
\epsfxsize=11.0cm 
\epsfbox{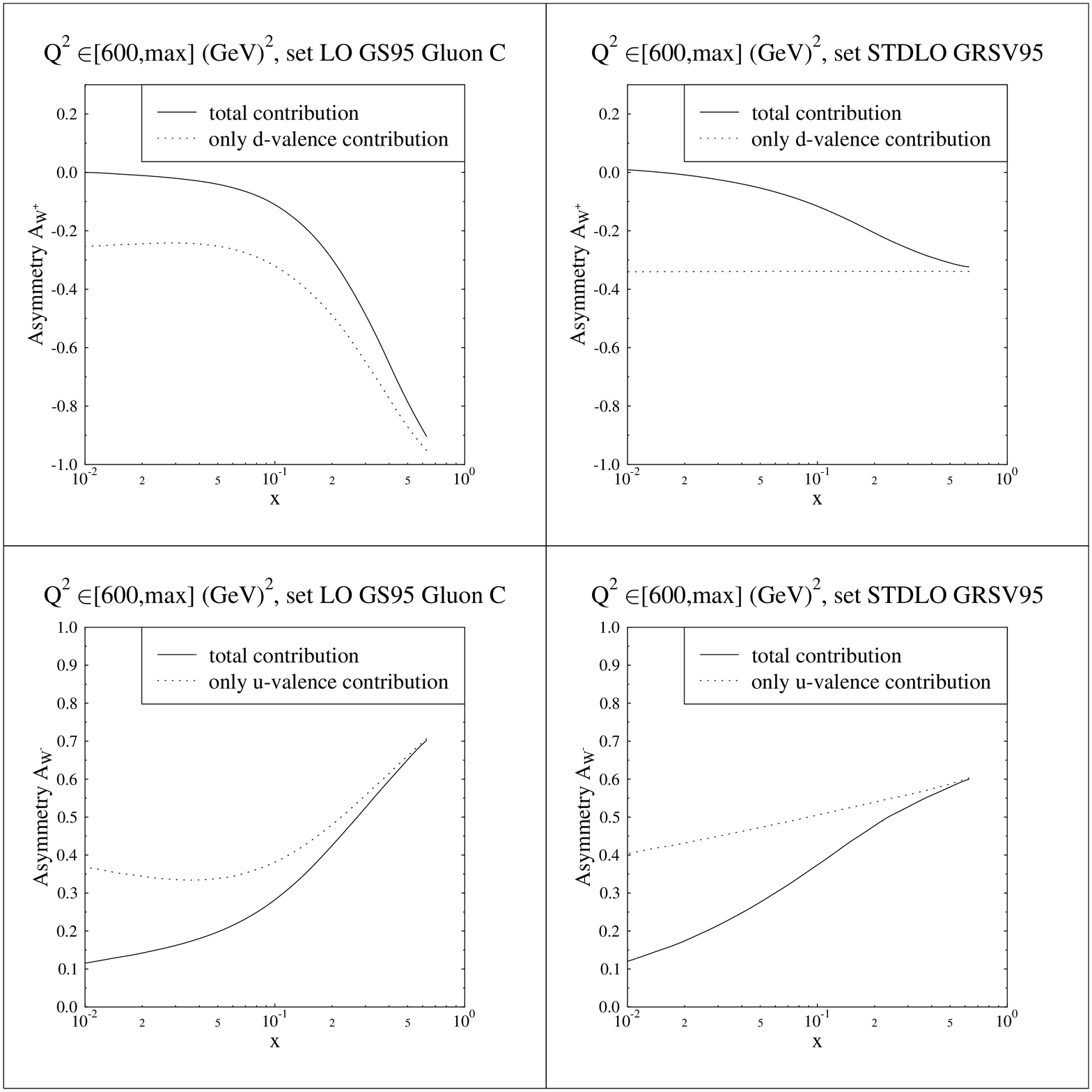}
\newline
\newline
\end{center}  
\begin{center}
{\bf Figure 2}
\end{center}
\newpage
\begin{center}
\leavevmode
\epsfxsize=11.0cm
\epsfbox{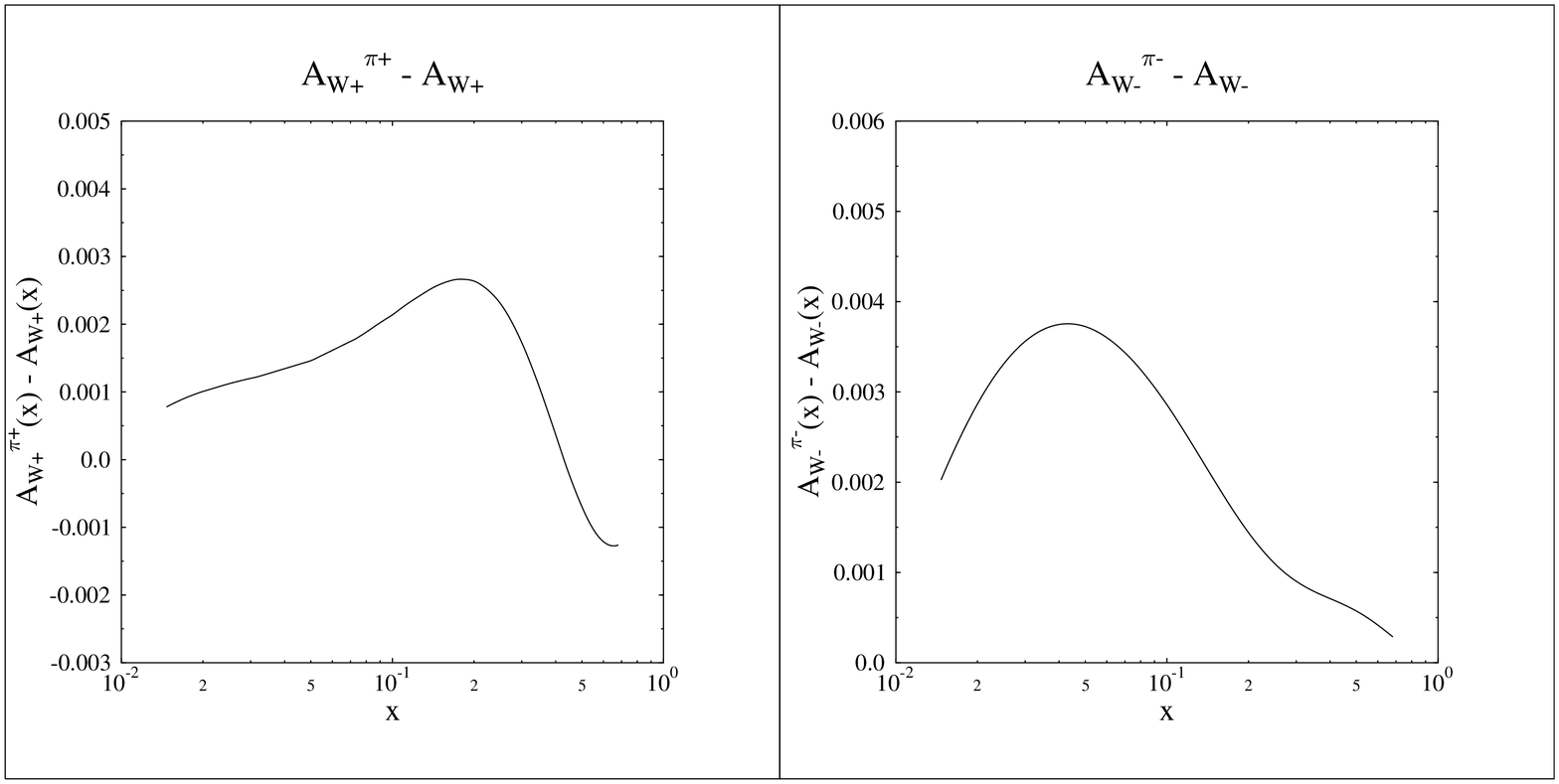}  
\newline
\newline
\end{center}
\begin{center}
{\bf Figure 3}
\end{center}
\begin{center}
\leavevmode
\epsfxsize=11.0cm
\epsfbox{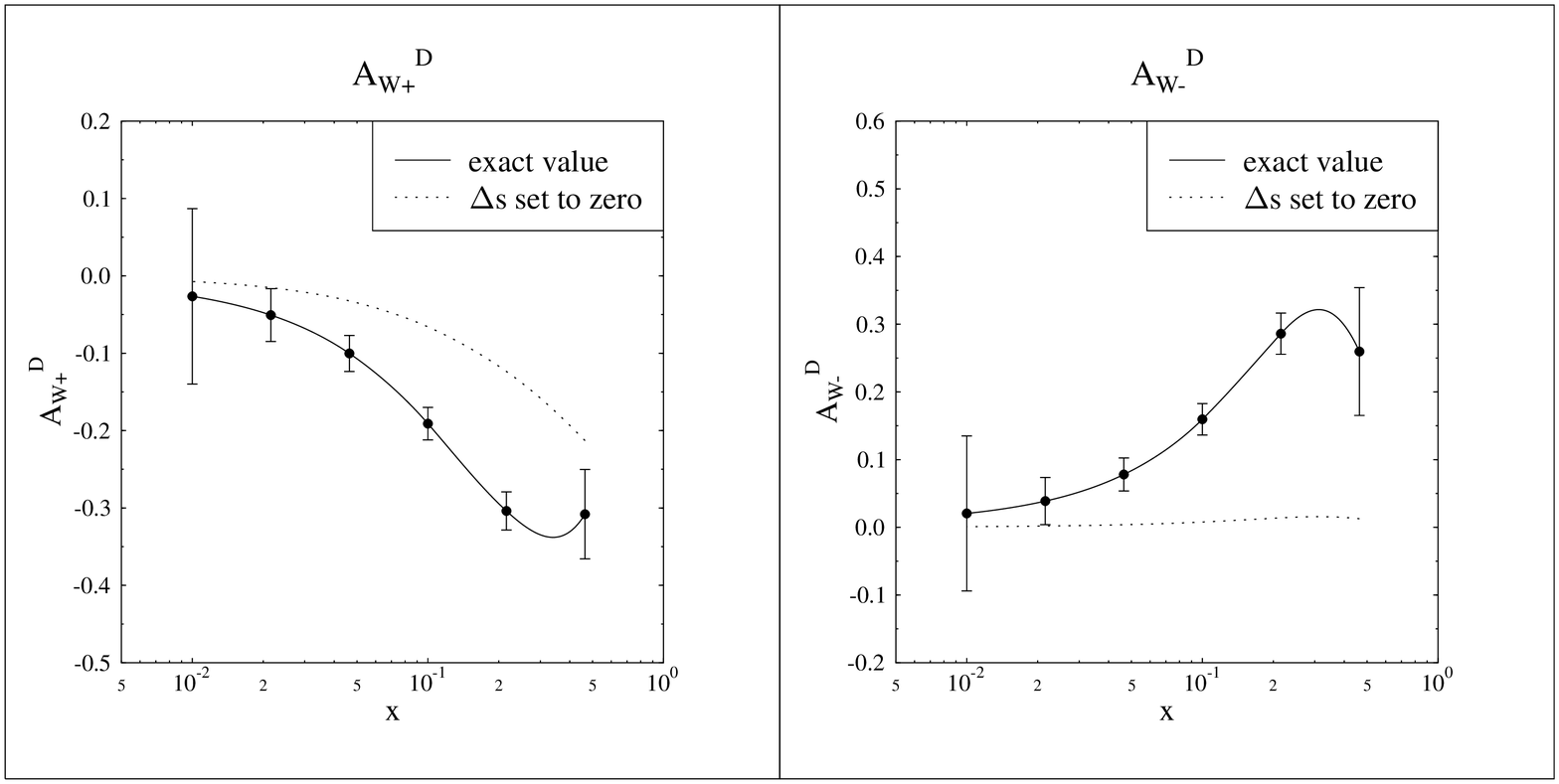}  
\newline
\newline
\end{center}
\begin{center}
{\bf Figure 4}
\end{center}
\end{document}